\documentclass{article}

\usepackage{arxiv}
\usepackage[utf8]{inputenc} 
\usepackage[T1]{fontenc}    
\usepackage{hyperref}       
\usepackage{url}            
\usepackage{booktabs}       
\usepackage{amsfonts}       
\usepackage{nicefrac}       
\usepackage{microtype}      

\usepackage{graphicx}
\usepackage{natbib}
\usepackage{doi}
\usepackage{amsmath}

\title{Large Impact of Small Vertex Cuts on the Mechanics of Origami Bellows}

\author{Mengzhu Yang \\
	Department of Aerospace Engineering \\University of Bristol Bristol\\BS8 1TR United Kingdom\\
	\texttt{mengzhu.yang@bristol.ac.uk} \\
	\And
	{Steven W. Grey} \\
	B2Space\\
	Newport, United Kingdom\\
	\texttt{steven.grey@b2-space.com} \\
	\And
    {Fabrizio Scarpa} \\
	Department of Aerospace Engineering \\University of Bristol Bristol\\BS8 1TR United Kingdom\\
	\texttt{F.Scarpa@bristol.ac.uk} \\
	\And
    {Mark Schenk}\thanks{m.schenk@bristol.ac.uk}\\
	Department of Aerospace Engineering \\University of Bristol Bristol\\BS8 1TR United Kingdom\\
	\texttt{m.schenk@bristol.ac.uk} \\
}

\renewcommand{\headeright}{}
\renewcommand{\undertitle}{}
\renewcommand{\shorttitle}{Large Impact of Small Vertex Cuts on the Mechanics of Origami Bellows}

\hypersetup{
pdftitle={Large Impact of Small Vertex Cuts on the Mechanics of Origami Bellows},
pdfsubject={},
pdfauthor={Mengzhu Yang, Steven W. Grey, Fabrizio Scarpa, Mark Schenk},
pdfkeywords={origami, origami bellows, non-rigid foldable, origami modelling},
}

\begin{document}
\maketitle

\begin{abstract}
 For origami structures, perforating or cutting slits along creases is an effective method to define fold lines and alleviate stress concentrations at vertices.
    In this letter we show numerically and experimentally that for non-rigid-foldable origami bellows (\textit{e.g.}\ Miura-ori, Kresling patterns) the introduction of small cut-outs at the vertices results in up to an order of magnitude reduction of the bellows' stiffness under axial compression. Further, the cut-outs at vertices impact the nonlinear response, \textit{e.g.}\ the position and magnitude of a force limit point and presence of bistable configurations. 
    As the origami bellows are not rigid foldable, an axial compression will necessarily result in facet deformations; the small cut-outs at the vertices are found to provide an unexpectedly large stress alleviation, resulting in disproportionate changes in mechanical properties of the bellows. In order to accurately model the mechanics of origami bellows, such manufacturing details must therefore be captured accurately.
    Lastly, introducing vertex cut-outs can be offered as a novel approach to tailoring the stiffness of non-rigid foldable origami structures.
\end{abstract}


\section{Introduction}
%
%
\noindent Origami, a traditional paper folding art, is a mapping of a flat sheet material to a three-dimensional object by folding along defined creases. Due to its inherent kinematics, programmable stiffness and self-foldability, origami has become an active research topic in fields such as deployable structures~\citep{wang2022design, dang2022inverse}, robotics~\citep{balkcom2008robotic,jeong2018design}, architected materials~\citep{silverberg2014using,li2019architected} and energy absorption devices~\citep{ma2011thin,ming2021energy}.
Origami bellows are formed by folding origami patterns into closed cylindrical structures. The bellows can be classified according to their crease patterns, including the well-known Miura-ori and Kresling patterns~\citep{reid2017geometry}. Origami bellows have been shown to provide a rich and tailorable nonlinear response, and have been explored for applications such as inflatable booms \citep{schenk2013geometry}, binary switches \citep{masana2020origami}, mechanical memory\citep{jules2022delicate}, and soft robotics \citep{pagano2017crawling,chen2020programmably,wu2021stretchable}.


For many applications, the mechanical properties of the origami bellows must be predicted accurately to achieve the desired functionality. Although multi-stability of the bellows can be predicted geometrically~\citep{reid2017geometry}, most bellows are not rigid-foldable~\citep{tachi2009simulation} and the facet deformations must thus be captured when modelling their folding motion.
A common simplifying assumption is that fold lines remain straight, allowing the origami bellows to be modelled analytically and numerically using axial bar elements placed along the fold lines of triangulated patterns~\citep{guest1996folding, zhai2018origami}. A revised model of the Kresling bellows also accounts for kinking of the folds observed in experiments~\citep{masana2019equilibria}. More general bar-and-hinge models simulate in-plane stretching using bar elements while capturing folding and out-of-plane bending of facets using rotational hinges~\citep{schenk2011origami, filipov2017bar, liu2018highly}. This method has been shown to be effective in simulating Miura-ori sheets~\citep{liu2017nonlinear}, Miura–ori tubes~\citep{grey2019strain}, \textit{etc}. 
In order to capture more detail and increase modelling fidelity the Finite Element Method (FEM) is commonly used~\citep{moshtaghzadeh2022prediction, shen2022experimental}. Various researchers also focused on the modelling of fold lines. For instance, Shen et al.~\citep{shen2021deployment} replaced the crease area by hinge connectors with an equivalent constitutive model based on numerical simulations, whereas Jules et al.~\citep{jules2019local} formulated analytical expressions for fold stiffness based on experiments. Hernandez et al.~\citep{hernandez2016modeling} introduced smooth folds in origami structures, allowing for the analysis of origami structures using shell representations for the folds.


Origami engineers and researchers may introduce holes at vertices to avoid large stress concentrations or perforate the fold lines to define a crease; their effects on the mechanics of origami structures have also been investigated. For rigid-foldable origami such as the Tachi-Miura Polyhedron~\citep{yasuda2013folding} and the Miura-ori tube~\citep{Grey2021embeddedactuation}, small vertex cuts show minimal influence on the structural stiffness. In contrast, for non-rigid origami, the impact of cuts may be significant. For instance, Yu et al.~\citep{yu2021cutting} identified that cutting a hole in a folded elastic disk could significantly impact the mechanical response, for example eliminating bistability. Hwang~\citep{hwang2021effects} studied the impact of offset slots along the crease line on the mechanical properties of Kresling pattern origami bellows. They found that the length and spacing of the slots were significant factors in the nonlinear response, including presence of instability.

\begin{figure}[htp]
 	\centering
 	    \includegraphics[width=0.8\linewidth]{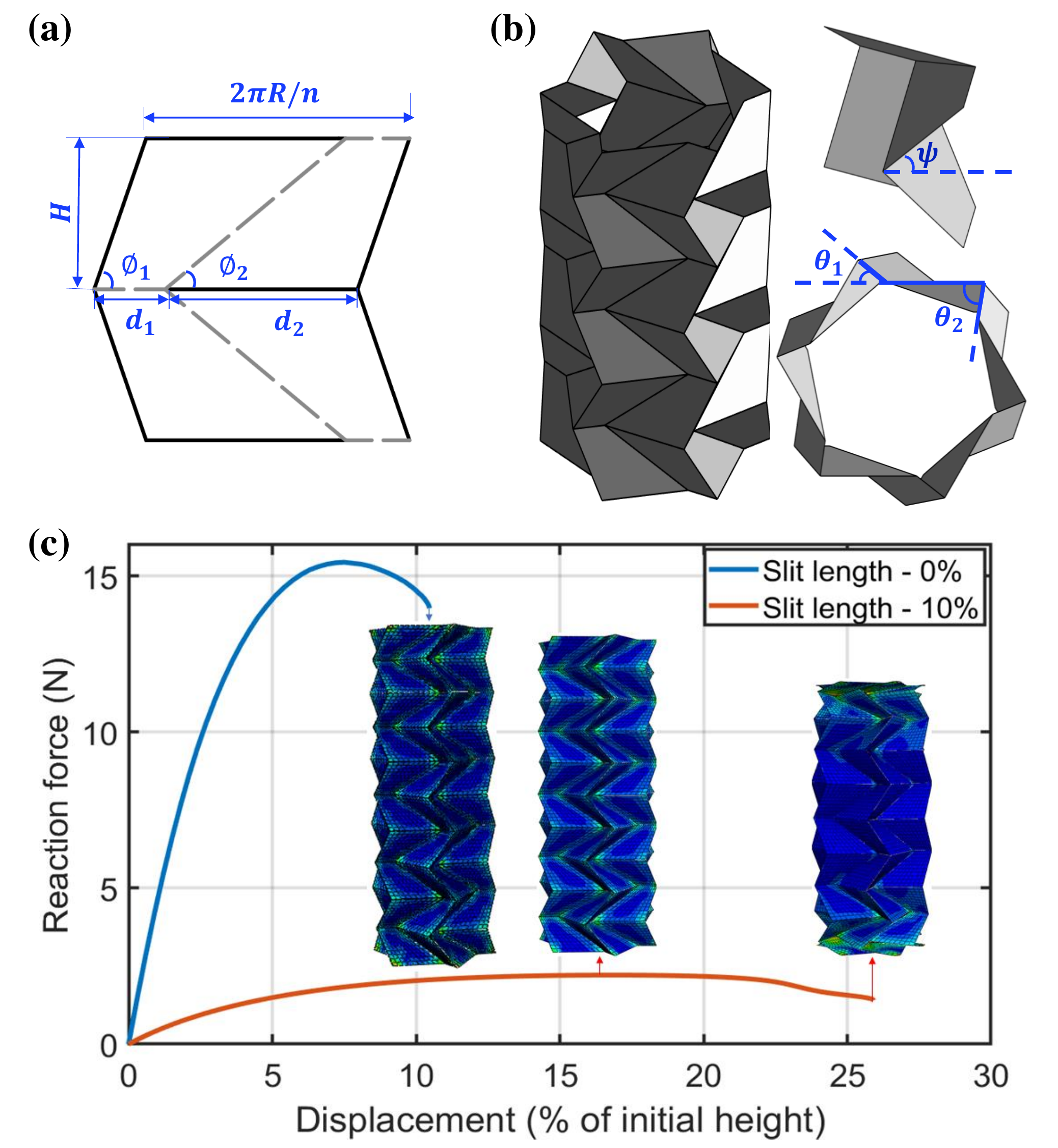}
        \caption{Geometry and mechanical response of a Miura-ori bellow. (a) Crease pattern of a Miura-ori unit cell, with sector angles $\phi_1$ and $\phi_2$, crease lengths $d_1$ and $d_2$, bellow radius $R$, facet height $H$ and where $n$, $m$ represent the number of Miura-ori unit cells around the circumference and along the length of the bellows respectively; note that for $d_1 = 0$ the Kresling pattern is recovered.
        %
        %
        (b) Assembled Miura-ori bellow, where $2\psi$ is the dihedral fold angle between adjacent facets and $\theta_1$ and $\theta_2$ are angles between two adjacent folds along the circumference.
        (c) Characteristic force-displacement responses of a Miura-ori bellow ($m=5$, $n=6$, $\phi_1=71^{\circ}$, $\phi_2=41^{\circ}$, $H=39$~mm, $R=65$~mm; bistable, flat-foldable) with slit length = 0\% and 10\% of the fold line length under axial compression. The introduction of a small vertex cut results in a decrease of over 90\% for the initial axial stiffness, as well as a large reduction in the force limit point.}
 	\label{fig:bellowsintro}
\end{figure}

For non-rigid foldable origami such as origami bellows, detailed numerical models have thus been found to be essential since small changes and imperfections may significantly influence the mechanical properties \citep{ming2021energy,yu2021cutting,hwang2021effects}. In our work, we introduce small cuts (or slits) at the vertices and study the impact on the mechanics of origami bellows. As an illustrative example, consider the response of the Miura-ori bellows shown in Fig.~\ref{fig:bellowsintro}.
The loading boundary condition constrains the top/bottom ends to remain planar, but allows the cross-section to expand or contract during compression. As shown in Fig.~\ref{fig:bellowsintro}(c), for the bellow without vertex cuts the response is nonlinear and the deformation is nominally uniform along the length of the bellow. Introducing slits at the vertices (slit length = $10\%$ of fold line) results in significant changes in response. Initially the deformation is uniform along the bellows, but this localises at the boundaries after reaching the load limit point. Moreover, the axial stiffness shows a decrease of over 90\%. This preliminary result shows an unexpectedly large impact of small slits at the vertices; a total reduction of 20\% of the fold line length results in a reduction of 90\% in axial stiffness. Therefore, it is essential to investigate the influence of the vertex cuts on the mechanical response of origami bellows.

The remainder of the paper is organised as follows. Section~\ref{sec:Numerical method and experimental setup} describes the numerical models of Miura-ori and Kresling bellows as well as the experimental setup. Results of simulations and experiments are analysed in Section~\ref{sec:Results and Discussion}. Findings in this letter are concluded in Section~\ref{sec:Conclusions}.

\section{Numerical and Experimental Methods} 
\label{sec:Numerical method and experimental setup}

Our aim is to model the axial compression of origami bellows. In addition to the pattern geometry and material properties, the bellows' full mechanical response depends on the loading boundary conditions in combination with the number of layers, $m$, as evidenced by the localisation of deformations observed in Fig.~\ref{fig:bellowsintro}(c). Under the selected boundary conditions, the initial deformation is nominally uniform along the length of the bellows; for simplicity we therefore study a single layer ($m=1$) to reveal the impact of the vertex slit length on the mechanical response of the bellows.

\subsection{Numerical Method}
\label{sec:Numerical method}
The origami bellows are modelled using the commercial finite element software Abaqus \cite{systemes2018abaqus} in order to capture the details of the vertex cuts. The facets of Miura-ori bellows are meshed using quadrilateral shell elements (S4R) while triangular shell elements (S3R) are utilized for Kresling bellows. All simulations are geometrically nonlinear. After a mesh sensitivity study (Appendix~\ref{AP:sensitivity-study}) a mesh size of $40 \times 40$ elements per facet was selected. Hinge connectors (CONN3D2 elements) with zero width and constant torsional stiffness are used to connect pairs of nodes on adjacent facets to represent the fold stiffness. Adjacent to each vertex, several pairs of mesh nodes along the fold lines remain unconnected to represent the vertex cuts; see Fig.~\ref{fig:Numerical model and experiment setup}(a). The size of the vertex cuts is quantified as \% of the fold line length. For bellows without slits, join connectors are used to connect the vertex at two adjacent facets.

Thermoplastic Polyethylene Terephthalate (PET) film is selected as the material for the bellows; the Young’s modulus, Poisson’s ratio and thickness are $E = 3200$~MPa, $~\nu$ = 0.43, and $t = 0.2$~mm, respectively. The fold stiffness per unit length of fold is taken to be $k_p = 0.05$~N/rad based on experimental data~\citep{grey2019strain,Grey2021embeddedactuation,grey2021embedded}. The fold stiffness depends on material properties as well as the crease type and manufacturing technique (\textit{e.g.}\ perforation, scoring, hemming). However, our study shows that the compressive stiffness of the bellows is not sensitive to changes in $k_p$ (Appendix~\ref{AP:sensitivity-study}). Note that the total stiffness along each fold line in the numerical models is kept constant, by increasing the fold stiffness $k_p$ proportionally. In folding a bellow from a flat sheet, self-stress is introduced at the fold lines, which may in turn affect its mechanical properties \citep{grey2020mechanics}. The choice of PET has the additional benefit of being able to anneal the material to relax initial stresses at the fold lines~\cite{sargent2019heat}. In our modelling, all fold lines are assumed to be stress free in the initial configuration.

To be representative of one layer in a multi-layer bellow, the loading boundary conditions are selected to allow free expansion/contraction of the bellow's cross-section during axial compression. The top and bottom of the layer are further constrained to remain planar, and a compressive displacement of 30\% of total height of the bellow is applied along the bellow's cylindrical axis. All other degrees of freedom on the top and bottom cross-sections are left unconstrained. This boundary condition will be referred to as ``Flexible BC''. In replicating the experimental results, the boundary conditions are changed to fix all translational degrees of freedom at the top/bottom cross-sections of the bellows; we refer to this as ``Fixed BC''.

\subsection{Experimental Method}
\label{sec:Experimental setup}
\begin{figure}
 	\centering
    \includegraphics[width=\linewidth]{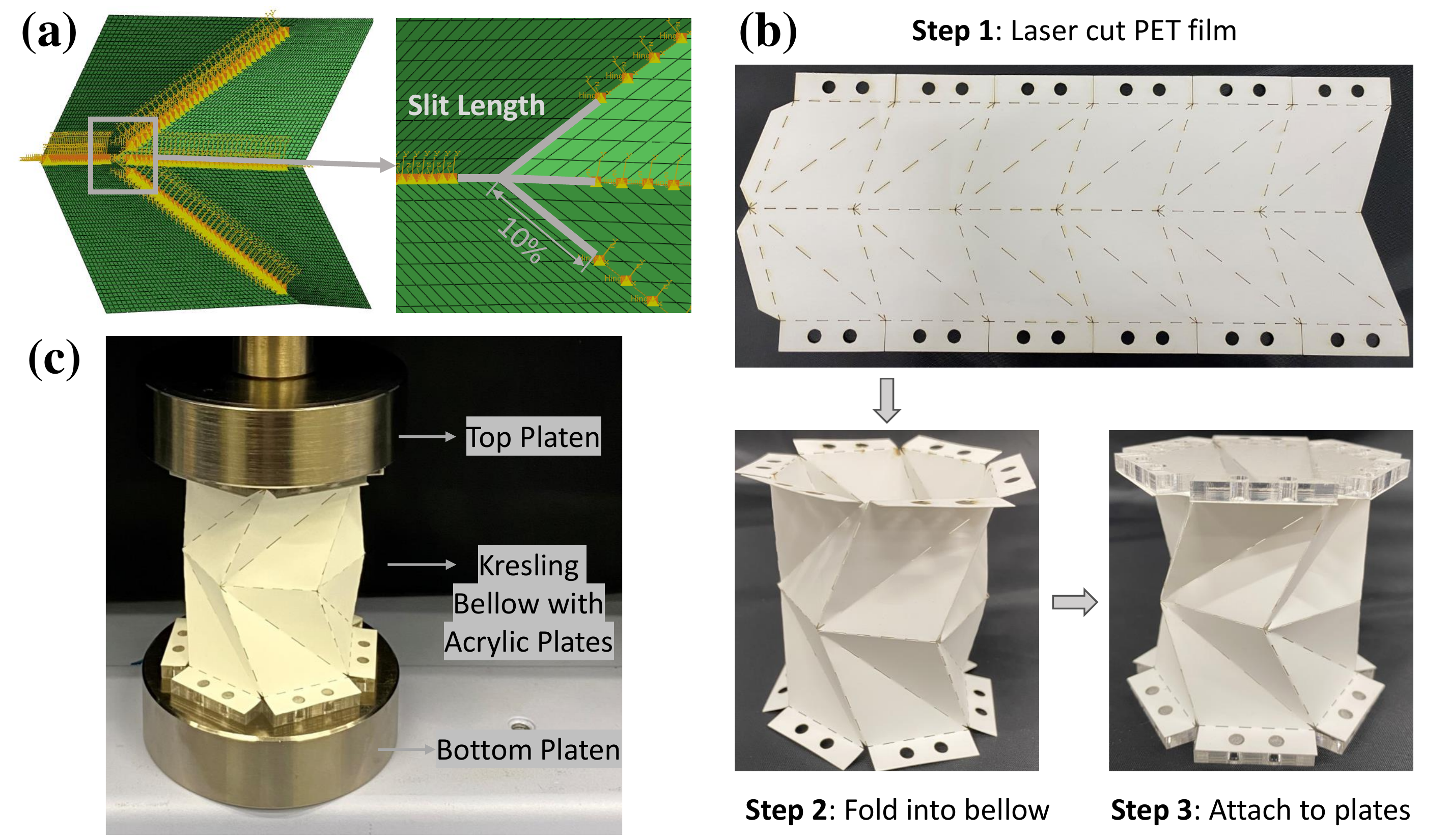}
    \caption{Numerical model, folding procedure and experiment setup for axial compression. (a) Distribution of hinge connectors along the fold lines of a unit cell of Miura-ori bellows; the mesh size is $40 \times 40$ elements per facet and 10\% of the fold line is cut open at the vertex, which is highlighted in grey. (b) Folding procedure of origami bellows from planar state to initial stable state. (c) Axial compression test setup for origami bellows.}
 	\label{fig:Numerical model and experiment setup}
\end{figure}

The numerical simulations of the Miura-ori and Kresling bellows are compared against experimental measurements, for samples with vertex cuts of 0\%, 5\%, 10\% and 20\% of the fold line length. The bellows are folded from PET sheets (PicoFilm Tearproof Colour Laser Film) with a thickness of $t = 0.2$ mm. The crease patterns are prepared using a Trotec Speedy 360 laser cutter. First, the pattern is cut out and fold lines are perforated with cuts of length CT = 1/9 of the fold line length (excluding slit length) spaced evenly at CT intervals to define the creases; see Fig.~\ref{fig:Numerical model and experiment setup}(b). For a slit length = 0\% of the fold line there is no perforation at the vertices. Next, the crease pattern is folded to form a single-layer bellow, which is closed and sealed using double-sided tape on the end tabs. If the bellow is flat-foldable, it is compressed to the flat state and allowed to spring back to the initial stable configuration. Finally, the top and bottom of the bellow are attached to acrylic plates using double-sided tape and bolts (for Miura-ori bellows). The experimental set-up thus constrains all displacements at the top and bottom cross-section of the bellows, whilst allowing folding along the fold lines. 
Once assembled, the bellows are compressed using a Shimadzu universal testing machine with 1 kN load cell at a speed of 2 mm/min; see Fig.~\ref{fig:Numerical model and experiment setup}(c). Three samples are manufactured and tested for each vertex cut size, and the averaged results are presented in Fig.~\ref{fig:Experimental validation}.

\section{Results and Discussion}
\label{sec:Results and Discussion}
\subsection{Experimental Validation of Numerical Models}
\label{sec:Experimental validation of numerical models}
\begin{figure}
 	\centering
 	    \includegraphics[width=\linewidth]{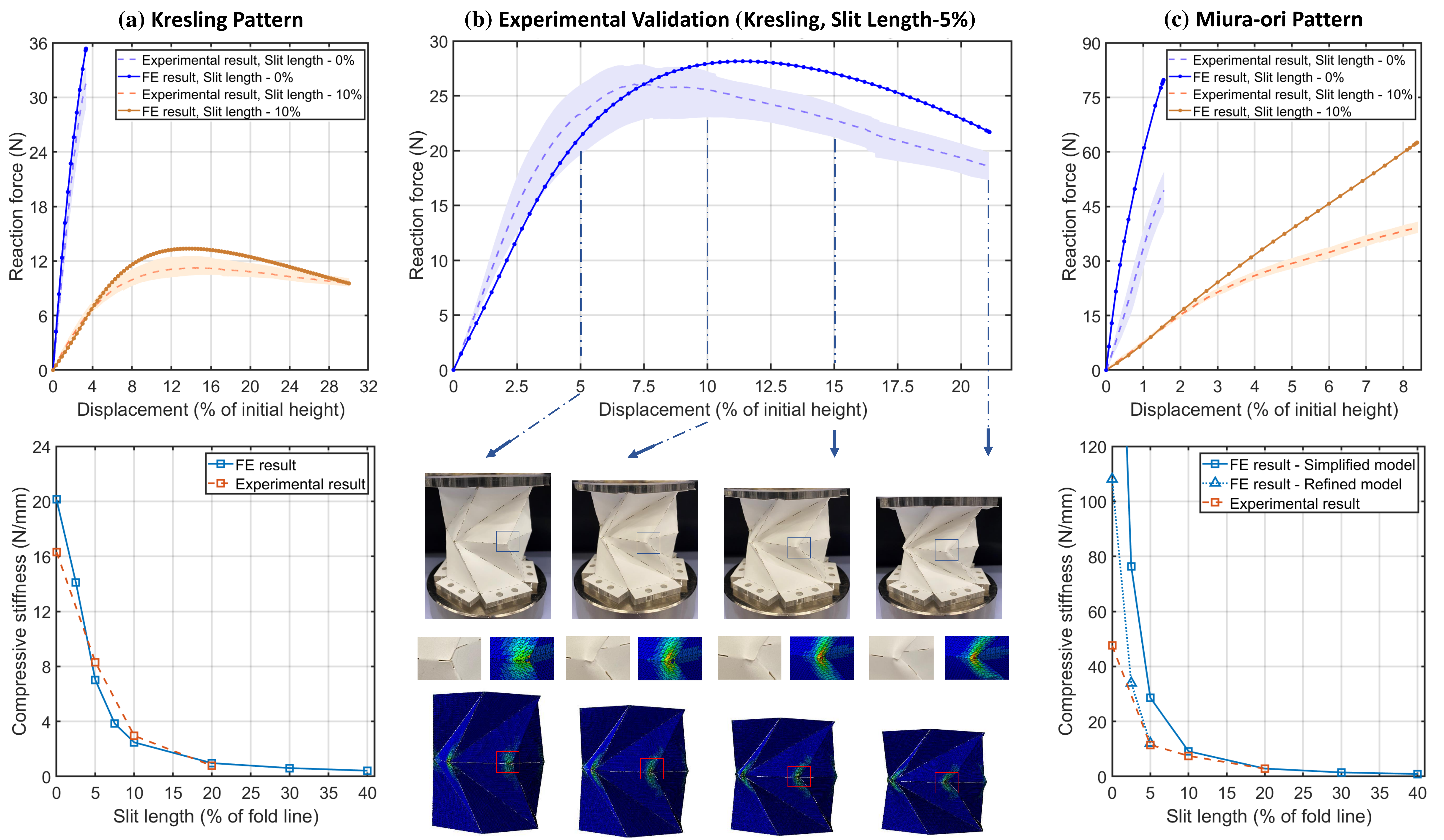}
        \caption{Experimental validation of numerical simulations (Fixed BC). (a) Kresling pattern ($n=6$, $\phi_1=71^{\circ}$, $\phi_2=38^{\circ}$, $H=37.27$~mm, $R=35$~mm, monostable). Top graph shows force-displacement curves for slit length = 0\% and 10\% of fold line. The shaded areas show the results of multiple experiments, dashed lines represent the average experimental results, and solid lines are simulations; the simulation results are in good agreement with experimental data. Bottom graph shows compressive stiffness as a function of slit length; the results demonstrate that introducing slits along the fold lines will lead to a significant decrease in axial stiffness of the bellows. (b) Kresling pattern with slit length = 5\% of the fold line. Qualitative comparison of deformed configurations at multiple displacements shows a good match between experiments and simulations, with both showing localised kinking of the folds at the vertices in the middle of the bellows. The simulation results show the von Mises stress, with warmer colours indicating higher values. (c) Miura-ori pattern ($n=6$, $\phi_1=71^{\circ}$, $\phi_2=41^{\circ}$, $H=39$~mm and $R=65$~mm, bistable). Top graph shows force-displacement curves for slit length = 0\% (refined FE model) and 10\% (simplified FE model) of fold line. Bottom graph shows compressive stiffness; numerical results for slit length = 0\%, 2.5\% and 5\% using a refined FE model are also included.
        }
 	\label{fig:Experimental validation}
\end{figure}

The numerical and experimental results for Kresling and Miura-ori bellows with different vertex slit lengths are presented in Fig.~\ref{fig:Experimental validation}. It is evident that the vertex cuts significantly impact the nonlinear response of the origami bellows; in particular, we highlight the  sensitivity of the bellows' compressive stiffness to the introduction of a small vertex cut. The compressive stiffness is calculated using a least-squares fit over displacements from $0-0.3$\% of the initial height. 

The experimental and numerical results for the Kresling bellows match well, and the relative error for the compressive stiffness is within 20\% (Appendix~ \ref{AP:Experimental validations}). The discrepancy between the simulation and the experiment can be attributed to factors such as self-weight, self-stress and imperfections which were not modelled; further, the slits along the fold lines at the top and the bottom of the bellows where the boundary condition is applied are not captured. Moreover, the deformations also show a good qualitative match. Figure~\ref{fig:Experimental validation}(b) shows the deformed configurations for a Kresling bellow with a slit length = 5\%. The experimentally observed localised kinking of the fold lines near the vertices is accurately captured in the numerical models.

\begin{figure}
   \centering
 	\includegraphics[width=\linewidth]{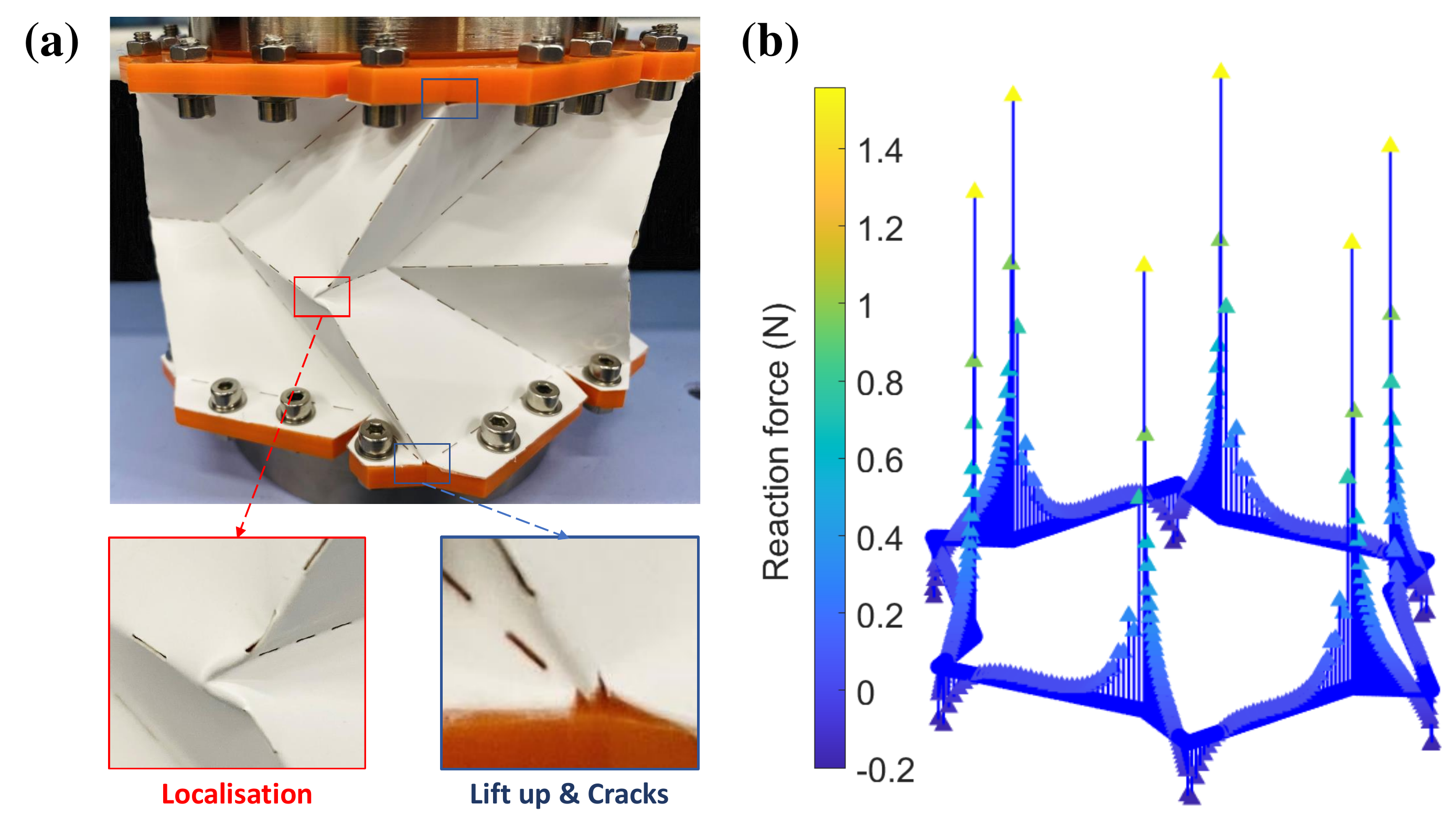}
    \caption{Details of deformed configuration and reaction force distribution of the Miura-ori bellows. (a) Deformed configuration in experiment at the end of compression, for slit length = 0\% of fold line; localised kinking of the folds at vertices in the middle of bellow (highlighted in red), and a ``lifting up'' effect and crack propagation (highlighted in blue) at  boundary vertices. (b) Reaction force distribution along the bottom of the Miura-ori bellow with at displacement = 0.3\% of initial height (Fixed BC). A negative reaction force is observed at vertices on the mountain fold (\textit{i.e.}\ pulling away from the end plates), while the reaction force at vertices on the valley fold is positive.}
 	\label{fig:miuraexplanation}
\end{figure}

For the Miura-ori bellows, discrepancies in the compressive stiffness were found for small vertex cuts (up to 5\% of the fold length), with the experimental results being much more compliant than the numerical simulations. Upon closer inspection, as highlighted in Fig.~\ref{fig:miuraexplanation}(a), it was noticed that at certain vertices the bellows would ``lift up'' from the end plates, resulting in more compliant boundary conditions than those imposed in the simulations. This is the result of unexpected tensile reaction forces (\textit{i.e.}\ the bellows pulling away from the plates) at these vertices, despite the bellows being loaded in compression; see Fig.~\ref{fig:miuraexplanation}(b).
To capture this effect, a more refined FE model included the details of the end tabs and enabled ``lift up'' at these vertices (see Appendix~\ref{AP:modified model} for further details). As shown in Fig.~\ref{fig:Experimental validation}(c) the refined model provides a good match for slit length = 5\% (and greater) with a relative error within 20\%, but not for slit length = 0\% of the fold line, where the predicted stiffness is approximately twice the experimental value.
It is hypothesised that small tears and cracks (see Fig.~\ref{fig:miuraexplanation}(a)) generated during the manufacturing or folding process will propagate during the compression procedure, serving as slits, and thus leading to a further reduction in stiffness. The cracks are randomly distributed at several fold lines and are approximately $0.9-1$ mm ($2.1-2.4$\% of the fold line) in length; the measured compressive stiffness for slit length = 0\% of the fold line falls between the numerical results of slit length = 0\% and 2.5\% of the fold line (refined FE model), which supports this assumption.

Overall, the numerical results show a good match with experimental results, allowing us to explore the nonlinear response of origami bellows further.

\subsection{Effect of Vertex Slit Length}
\label{sec:Effect of slit length on compressive stiffness}
\begin{figure}
 	\centering
 	\includegraphics[width=\linewidth]{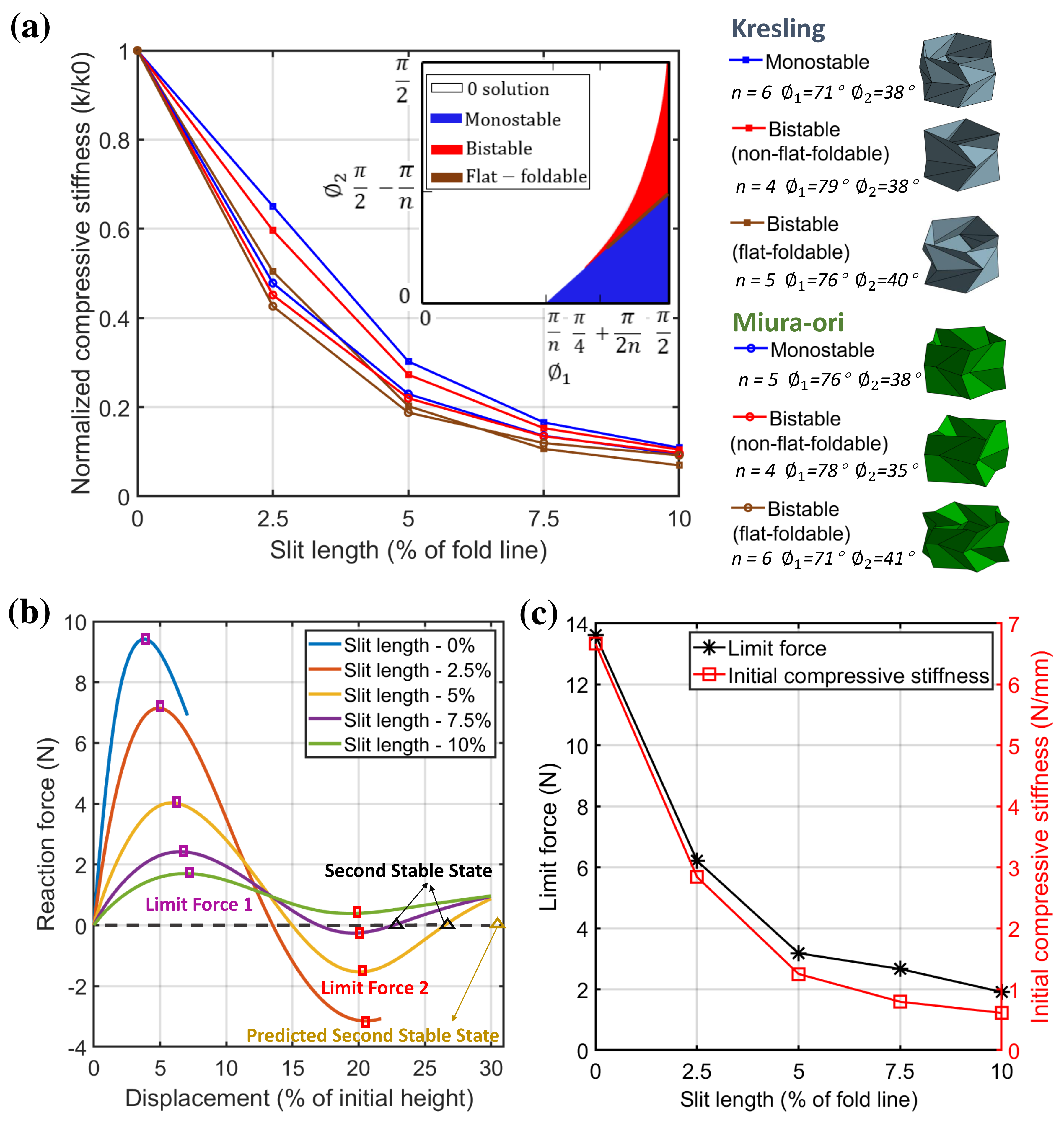}
    \caption{Numerical simulations of the mechanics of origami bellows (Flexible BC). (a) Normalised compressive stiffness of Miura-ori and Kresling pattern FE models with different geometric parameters, stability and slit length. Inset is the geometric parameter space of Miura-ori and Kresling bellows~\citep{reid2017geometry}; the monostable region, bistable region and flat-foldable region are indicated. All of the curves show significant decrease in compressive stiffness with the increase in slit length. (b) Force-displacement curves of a non-flat-foldable, bistable Kresling bellow ($n=4$, $\phi_1=79^{\circ}$, $\phi_2=38^{\circ}$, $H=40$~mm, $R=30.7$~mm) with varying slit length. The force limit points as well as the second stable state are highlighted. The figure shows the change of the magnitude and location of the force limit point. (c) Limit force \textit{vs} slit length curve and corresponding compressive stiffness \textit{vs} slit length curve of a flat-foldable, bistable Miura-ori bellow ($n=6$, $\phi_1=71^{\circ}$, $\phi_2=41^{\circ}$, $H=39$~mm, $R=65$~mm); the trend of both curves is the same.}
     \label{fig:simualtion flexible BC}
 \end{figure}

The experimental and numerical results show that the compressive stiffness of the Kresling and Miura-ori bellows decreases sharply when the vertex slit length increases from 0\% to 10\% of the fold line length. This result is found to hold for a wide range of Kresling and Miura-ori bellows geometries. Figure~\ref{fig:simualtion flexible BC}(a) shows the normalised axial stiffness \textit{vs} vertex slit length for bellows patterns selected from across the feasible geometric design space~\citep{reid2017geometry}. This includes origami bellows that are monostable, bistable flat-foldable and bistable non-flat-foldable. Introducing slits at vertices of merely 2.5\% of the fold line length will lead to an almost 50\% decrease in stiffness compared to models with no slits; increasing the cut-out to 10\% of the fold line length will lead to a stiffness reduction by an order of magnitude.

The introduction of vertex cuts not only affects the compressive stiffness of the origami bellows, but also their nonlinear response. The force-displacement response of a bistable, non-flat-foldable Kresling model is shown in Fig.~\ref{fig:simualtion flexible BC}(b). For increasing slit length, the magnitude of the force limit point decreases and the location of the second stable state changes. What is more, for a slit length of 10\% of the fold line, the second stable point disappears, thus eliminating the predicted bistability.
Miura-ori bellows display a similar change in magnitude and location of the force limit point. For example, the force limit point of a flat-foldable bistable Miura-ori bellow with varying vertex cut length is shown in Fig.~\ref{fig:simualtion flexible BC}(c), and it is compared against the corresponding initial compressive stiffness. The trends of both curves is the same: the limit force also shows a decrease of over an order of magnitude when the slit length increases to 10\% of the fold line.

In summary, the introduction of a small vertex cut has been found to have an unexpected and disproportionate effect on the mechanics of origami bellows. The reduction in stiffness of the bellows is much greater than could be explained simply by the reduction in fold line length (and thus total fold stiffness). This suggests that the effect is due to the fact that the bellows are not rigid-foldable, and an axial compression therefore introduces in-plane loads in the facets. This hypothesis is supported by the analysis of a rigid-foldable Miura-ori unit cell (Appendix~\ref{AP:planar-Miura-ori}) as well as previous work where vertex cut-outs were used in rigid-foldable Miura-ori tubes and sheets~\citep{grey2019strain,Grey2021embeddedactuation}. Nonetheless, the sensitivity to the vertex slit length remains unexpectedly large. Hwang~\citep{hwang2021effects} described a similar effect, with a significant change in stiffness and nonlinear response of Kresling bellows as a result of perforations along the fold line. In that work, however, the shear stiffness of the fold line is drastically reduced as a result of the perforation pattern, which could account for the change in mechanical properties of the bellows.

\section{Conclusions}
\label{sec:Conclusions}
%
In this letter we investigate the impact of introducing vertex cuts in non-rigid foldable origami bellows. The mechanical response of the bellows (\textit{e.g.}\ axial compressive stiffness, nonlinear response) is found to be highly sensitive to the size of the cut-outs; the numerical results were validated against experiments. This effect is robust to different bellows patterns such as the Miura-ori and Kresling, as well as modelling approaches for the fold lines such as torsion springs and smooth folds (the implementation is discussed in Appendix~\ref{AP:Implementation of Smooth Fold Line Model}). This phenomenon has not been reported in previous studies, and the impact of cut-outs is not present in rigid-foldable patterns such as Miura-ori sheets.

Specifically, increasing the vertex slit length results in a disproportionate decrease in compressive stiffness of the origami bellows. As the slit length increases from 0 to 2.5\% of the fold line, the compressive stiffness drops by over 50\%. Once the slit length increases to 10\%, the stiffness decreases by an order of magnitude compared to the original bellows. The magnitude and location of force limit points in the nonlinear response are similarly impacted by the size of the vertex cut. Moreover, the vertex cut also affects the second stable state and may lead to the disappearance of the predicted bistability. In the analysis of Miura-ori bellows, another interesting effect was observed: for a bellows under compression, a tensile reaction force was found at certain vertices, resulting in the bellows pulling away from the end plates at these locations. 


In conclusion, for non-rigid foldable origami bellows where folding also induces facet deformations, it is critical to accurately model the cut-outs at vertices, since it could lead to a significant change in mechanical properties, such as axial stiffness, limit force and stability. The importance of capturing particular details is also evidenced by recent work on perforated fold lines \citep{hwang2021effects}. These results may have wider implications for the applicability of reduced-order models (such as bar and hinge models), which cannot capture such details, for the analysis of non-rigid-foldable origami. On the other hand, tailoring the vertex slit length may provide an alternative method to manipulate stiffness and stability of origami bellows, and can therefore offer a promising approach to designing mechanical memory systems, graded-stiffness structures, \textit{etc}.

\section*{Acknowledgement}
Mengzhu Yang is supported through a China Scholarship Council (CSC)--University of Bristol joint scholarship.

\bibliographystyle{unsrtnat}
\bibliography{references}  

\clearpage
\newpage


\appendix
\section{Geometry of Miura-ori Bellows}
\label{AP:Geometry of Miura-ori bellow}
\noindent The crease pattern of Miura-ori bellows is illustrated in Fig.~\ref{fig:bellowsintro}(a) with the corresponding three-dimensional model shown in Fig.~\ref{fig:bellowsintro}(b). The pattern can be described by sector angles $\phi_1$ and $\phi_2$, vertex spacing $d_1$ and $d_2$, radius $R$, layer height $H$, number of unit cells around the circumference $n$ and number of layers along the length of the bellows $m$. For Miura-ori bellows $\phi_1$ and $\phi_2$ satisfy:
\begin{equation}
    0 < \phi_2  < \phi_1 < \frac{\pi}{2}
\end{equation}
and
\begin{equation}
    d_1+d_2=\frac{2\pi R}{n}
\end{equation}
The length parameters $d_1$ and the layer height $H$ are geometrically restricted to \citep{reid2017geometry}:
 \begin{equation}
   0 < H \leq \frac{\frac{2\pi R}{n}-d_1}{\cot\phi_2-\cot\phi_1}
 \end{equation}
If the bellow is flat-foldable, $H$ should satisfy:
 \begin{equation}
    H= \frac{\frac{2\pi R}{n}-2d_1}{\cot\phi_2-\cot\phi_1}
    \label{EQ3}
 \end{equation}
Following Reid et al.\ \citep{reid2017geometry}, cylindrical closure conditions on $\theta_1$ and $\theta_2$ (see Fig.~\ref{fig:bellowsintro}(b)) are introduced:
\begin{equation}
   \theta_2=\pi-\frac{2\pi}{n}-\theta_1
    \label{EQ1}
\end{equation}
and
%
\begin{equation}
  \tan{\frac{\theta_1}{2}}=\frac{1}{2\tan\frac{\pi}{n}}\left[1-\frac{\tan\phi_2}{\tan\phi_1} -\sqrt{\left(\frac{\tan\phi_2}{\tan\phi_1}-1\right)^2-4\frac{\tan\phi_2}{\tan\phi_1}\tan^2\frac{\pi}{n}}\right]
   \label{EQ2}
\end{equation}
%
from which angle $\psi$ (Fig.~\ref{fig:bellowsintro}(b)) can be calculated as:
\begin{equation}
   \sin\psi=\frac{\tan\frac{\theta_1}{2}}{\tan\phi_2}
\end{equation}
To satisfy Equations \ref{EQ1} and \ref{EQ2}, the design space of the bellows can be divided into three regions \citep{reid2017geometry}: (i) invalid parameters, (ii) mono-stable region (iii) and bi-stable region. This definition of multi-stability does not depend on strain energy considerations, but rather geometric compatibility.
According to above equations, design parameters can be simplified into $\phi_1$, $\phi_2$, $R$, $d_1$, $H$, $m$ and $n$ --- the remaining parameters can be derived. Moreover, when $d_1=0$, the Miura-ori pattern reduces to the Kresling pattern.

\clearpage
\newpage

\section{Impact of Slit Length on Miura-ori Sheet}
\label{AP:planar-Miura-ori}
\noindent In previous work on a rigid-foldable Miura-ori tube, the size of the vertex cut-out had not been found to impact its mechanical response \citep{grey2019strain}. This observation is here demonstrated for a single planar Miura-ori unit cell. 

\begin{figure}[h]
 	\centering
 	  \includegraphics[width=0.8\linewidth]{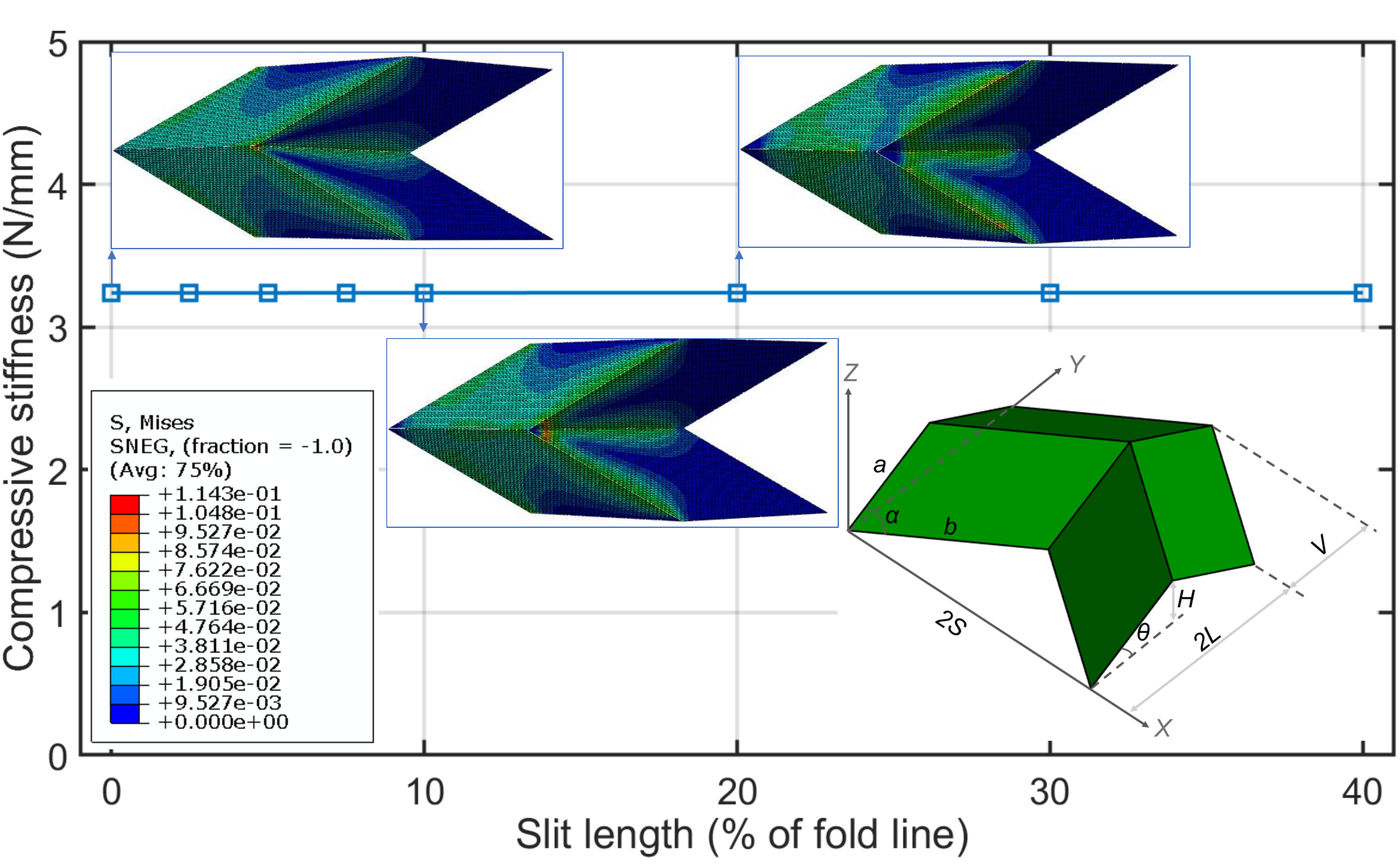}
    \caption{Compressive stiffness of a single unit cell in a regular Miura-ori sheet ($a=40$~mm, $b=40$~mm, $\alpha=40^{\circ}$, $\theta=45^{\circ}$) is insensitive to slit length. The inset is the geometry of a Miura-ori unit cell.}
   \label{fig:unitMiura}
\end{figure}

The unit cell geometry is shown inset in Figure~\ref{fig:unitMiura}. Dimensions $a$ and $b$ are crease lengths, $\alpha$ is facet angle, $\theta$ is the angle between the facets and the horizontal base; the remaining geometric parameters can be calculated \citep{schenk2013geometry}.
The material parameters are the same as those used in the main letter. The nodes in the $yz$-plane are constrained in the $x$-direction, while a compressive displacement of $30\% \times 2S$ is applied on the other end of the unit cell. 
%
%
All other degrees of freedom on both two ends are left unconstrained. The slit length changes from 0 to 40\% of the fold line, and the total fold stiffness along each fold line maintains the same.

Figure~\ref{fig:unitMiura} shows that the compressive stiffness of the unit cell remains nominally constant for increasing slit length at the vertex. This illustrates that for a rigid-foldable pattern, in contrast with the origami bellows, the effect of the slit length can in effect be neglected.

\clearpage
\newpage

\section{Sensitivity of Bellows Axial Stiffness}
\label{AP:sensitivity-study}
\noindent In the main text, the compressive stiffness of the origami bellows is shown to be highly sensitive to the slit length; here we show that this phenomenon is not limited to the particular combination of fold stiffness and material thickness used in the main letter. And we also demonstrate the minimal sensitivity to fold stiffness and mesh density.

\begin{figure}[ht]
\centering
 	    \includegraphics[width=0.98\linewidth]{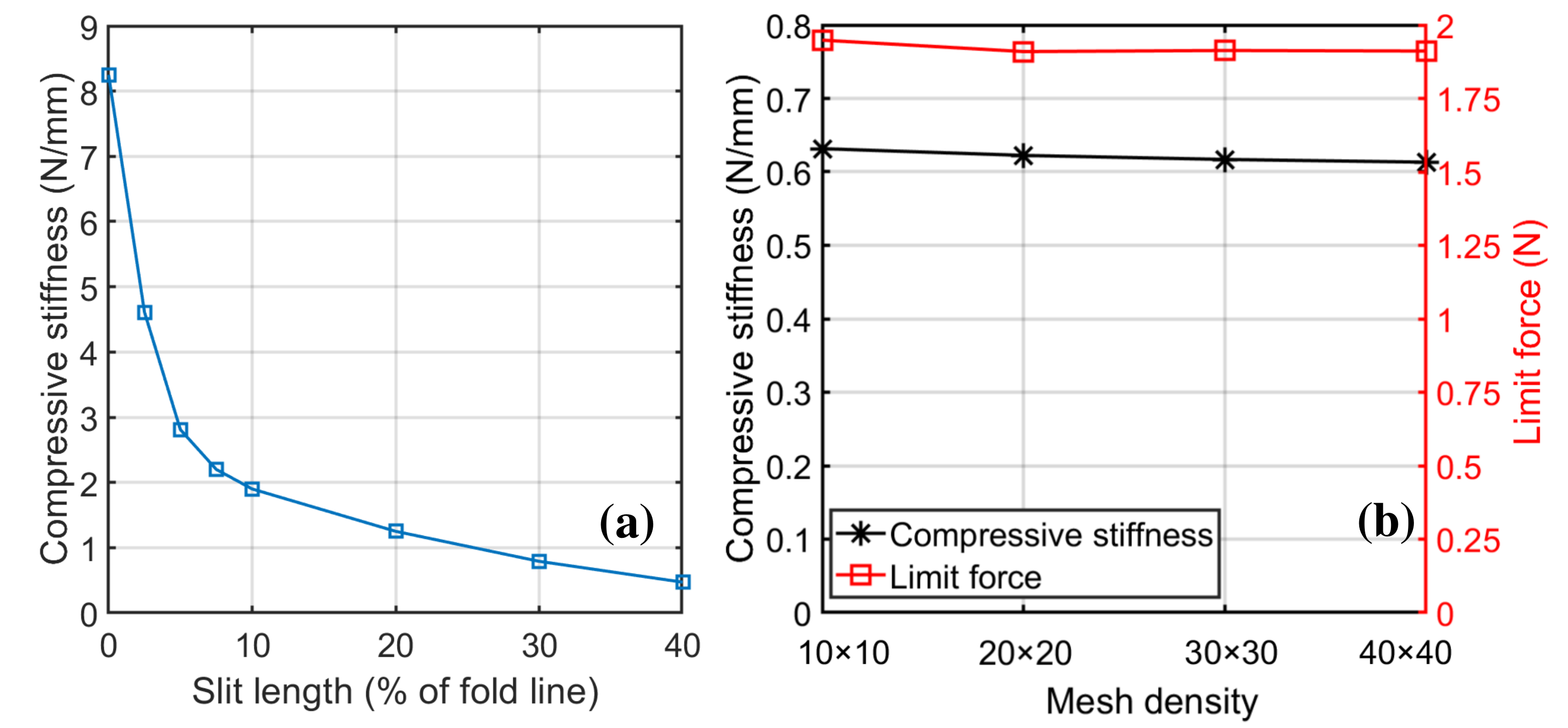}
 	    \label{fig:Sensitivity of compressive stiffness}
    \caption{Sensitivity of compressive stiffness to fold line stiffness and mesh density. (a) Compressive stiffness of Miura-Ori bellows with $k_p = 0.5$~N/rad in respect to different slit length; showing a huge impact of slit length. (b) Bellows' compressive stiffness modelled using hinge model with slit length equal to 10\% of fold line for increasing facet mesh density.}
   \label{fig:compressive stiffness sensitivity}
\end{figure}

First, we increase the fold stiffness per unit length of fold $k_p$ to 0.5 N/rad and vary the slit length from 0 to 40\%. The other parameters remain unchanged. The significant decrease in axial compressive stiffness with respect to slit length is again observed; see Fig.~\ref{fig:compressive stiffness sensitivity}(a).
Next, for the Miura-ori bellow with slit length = 2.5\% of the fold line, we increase the fold stiffness per unit length of fold by 5\%, 10\% and 15\%, namely $k_p = 0.05$, $0.0525$, $0.055$, $0.0575$ N/rad. The compressive stiffness $k_c$ increases 0.67\%, 1.23\% and 1.97\% correspondingly, see Table~\ref{tab:Sensitivity to Fold Line Stiffness}, which shows that the influence of $k_p$ on the compressive behaviour is not distinct.
Lastly, a mesh sensitivity study was conducted, increasing the mesh density from $10\times10$ to $40\times40$ elements per facet whilst keeping the slit length equal to 10\% of the fold line length. Both Fig.~\ref{fig:compressive stiffness sensitivity}(b) and Table~\ref{tab:Sensitivity to mesh density} show that if the mesh density is finer than $10\times10$, the compressive stiffness $k_c$ as well as limit force $F_l$ are almost the same (maximum difference is within 3\%) under different mesh sizes. However, in order to simulate fine slits such as slit length = 2.5\% of the fold line, mesh density is chosen as $40\times40$ elements per facet.

\begin{table}[ht]
    \caption{Sensitivity to Fold Line Stiffness, Slit Length - 2.5\%}
    \centering
    \label{tab:Sensitivity to Fold Line Stiffness}
    \setlength{\tabcolsep}{4.3mm}{
    \begin{tabular}{c c c c }
    \hline
        \textbf{$k_p$~(N/rad)} & \textbf{Increase~(\%)} & \textbf{$k_c$~(N/mm)} & \textbf{Increase~(\%)}  \\
    \hline
        0.05 & - & 2.841 & -  \\
        0.0525  & 5\% & 2.860 & 0.67\% \\ 
        0.055  & 10\% & 2.879 & 1.23\%  \\ 
        0.0575 & 15\% & 2.897 & 1.97\%  \\
    \hline
    \end{tabular}
    }
\end{table}

\begin{table}[ht]
    \caption{Sensitivity to Mesh Density, Slit Length - 10\%}
    \centering
    \label{tab:Sensitivity to mesh density}
    \begin{tabular}{c c c c c}
    \hline
        \textbf{Mesh density} &
        \textbf{$k_c$~(N/mm)} &
        \textbf{Change~(\%)} & \textbf{$F_l$~(N)} & \textbf{Change~(\%)}  \\
    \hline
        $10\times10$ & 0.632 & - &1.948 & -  \\
        $20\times20$   & 0.622 & -1.58\% & 1.910 & -1.95\% \\
        $30\times30$   & 0.617& -2.37\% &1.913 & -1.80\%  \\ 
       $40\times40$  & 0.614 & -2.85\% & 1.911 & -1.90\%  \\ 
       \hline
    \end{tabular}
\end{table}

\clearpage
\newpage

\section{Refined FE Model for Miura-ori}
\label{AP:modified model}
\noindent In experiments of Miura-ori bellows with slit length = 0\% and 5\% of fold line, debonding between plates and the ends of bellows is observed. This phenomenon is caused by the unique reaction force distribution at the bottom of the Miura-ori bellows; see Fig.~\ref{fig:miuraexplanation}(b). A positive reaction force is observed at the vertices at valley folds while reaction force at the vertices at mountain folds is negative (\textit{i.e.}\ pulling upwards from base, despite global compression load on bellows). Our study also shows that this phenomenon is robust to different compressive displacements and slit lengths. Therefore, vertices at mountain folds have the trend to pull away from the plates.

\begin{figure}[hb]
   \centering
    \includegraphics[width=0.85\linewidth]{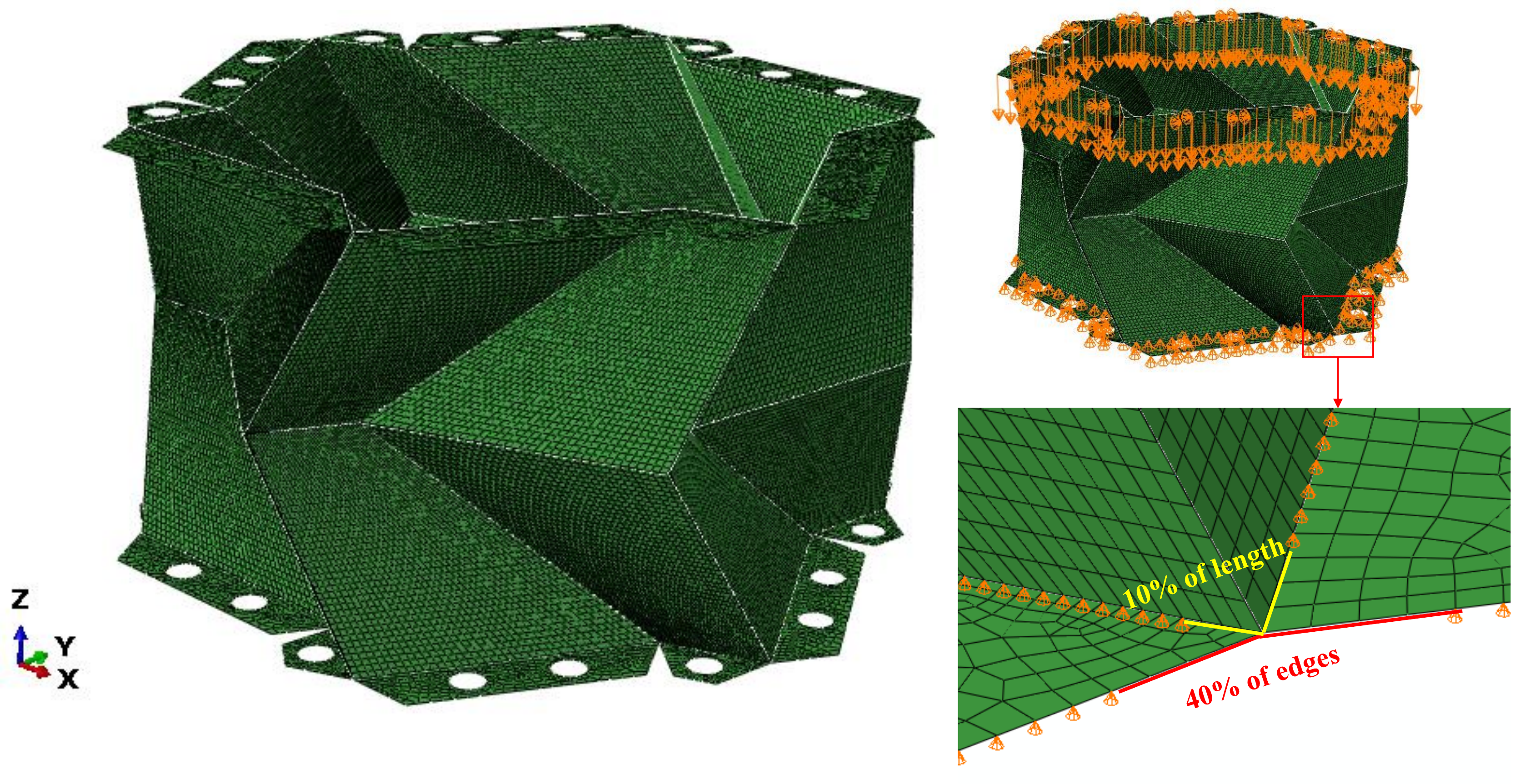}
    \caption{A refined numerical model capturing connecting tabs: 40\% of the edges of tabs ends and 10\% of the fold lines are left unconstrained around the vertices at the bottom and top to enable the vertex to pull up.}
 	\label{fig:modified numerical model}
\end{figure}

In order to capture the debonding effect, the numerical model was refined to reflect the experimental configuration in more detail; see Fig.~\ref{fig:modified numerical model}. At the vertices near the end plates, 40\% of the edges of the end tabs and 10\% of the fold lines are left unconstrained to facilitate the lifting up phenomenon; see Fig.~\ref{fig:modified numerical model}. For the remaining edges of the end tabs, the translational motion on the z-axis is constrained; all translational degrees of freedom are constrained at the holes. A compressive displacement of 30\% of the initial height of the bellow is applied along the bellows’ cylindrical z-axis at the edges of the end tabs and holes on the top.

\clearpage
\newpage

\section{Experimental Validation Miura-ori}
\label{AP:Experimental validations}
\begin{figure}[b]
    \centering
 	\includegraphics[width=0.9\linewidth]{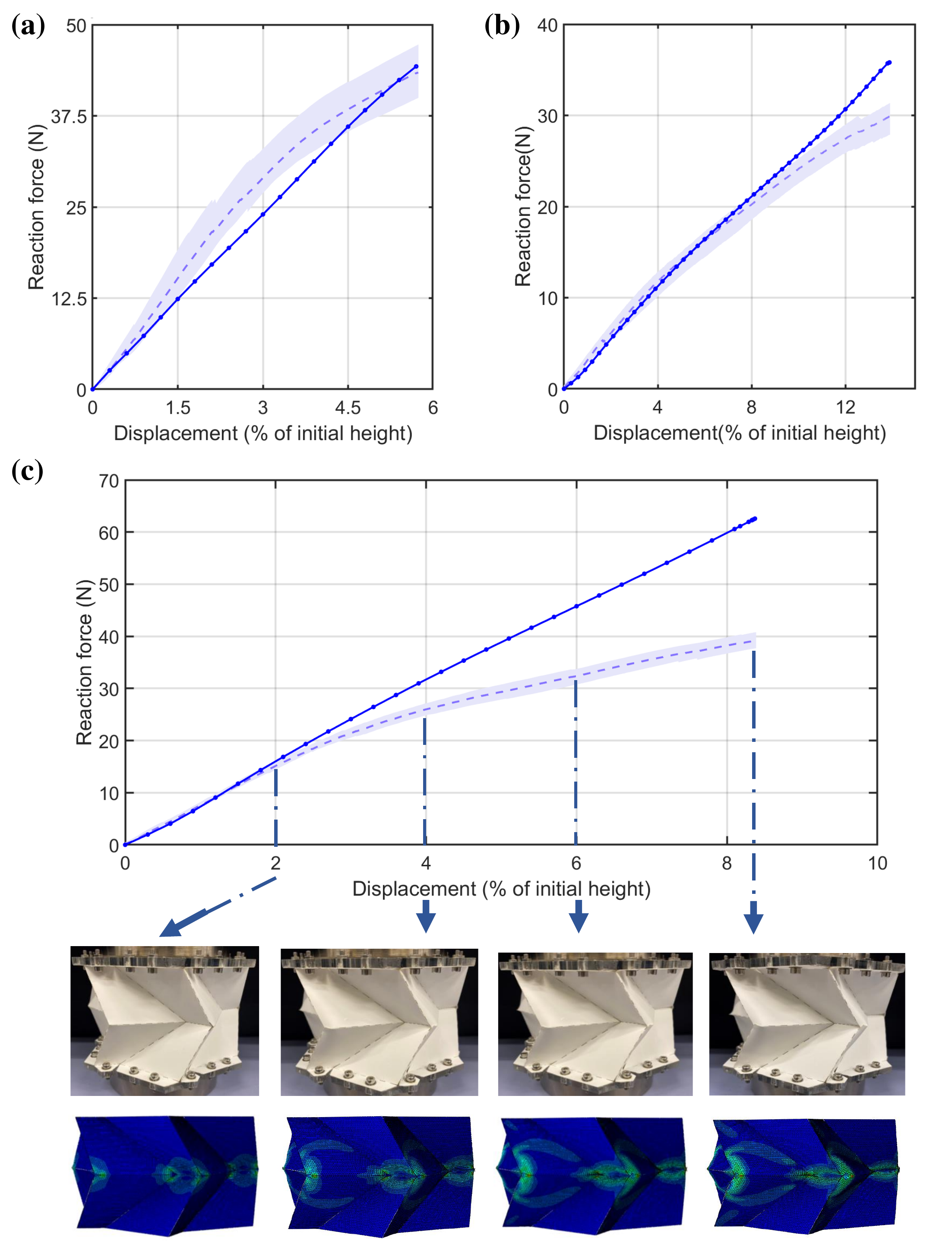}
    \caption{Experimental validations of numerical simulations of Miura-ori bellows ($n=6$, $\phi_1=71^{\circ}$, $\phi_2=41^{\circ}$, $H=39$~mm and $R=65$~mm, bistable) with different slit lengths. (a) Slit length = 5\% of fold line (refined FE model). (b) Slit length = 20\% of fold line  (simplified FE model). (c) Slit length = 10\% of fold line  (simplified FE model; inset are deformed configurations in experiment and simulation with corresponding Von Mises stress (warmer colours indicating higher values). The reaction force and deformation pattern shows good consistency, both showing localisation at the vertices in the middle of the bellow.}
 	\label{fig:Experimental validation of Kresling and Miura}
\end{figure}

\noindent This section presents further experimental validations of simulation results for Miura-ori bellows (results for Kresling are presented in the main text). Fig.~\ref{fig:Experimental validation of Kresling and Miura}(a)-(c) depict the experimental and numerical force-displacement responses of Miura-ori bellows with different slit lengths. For bellow with slit length = 5\%, 10\% and 20\%, the experimental results match well with the simulations, and the relative error for initial compressive stiffness is within 20\%. The compression of the Miura-ori bellow with slit length= 10\% is shown in Fig.~\ref{fig:Experimental validation of Kresling and Miura}(c) and the deformed configurations show good qualitative consistency. 

\begin{table}[h]
    \caption{Validation of compressive stiffness (N/mm) - Kresling}
    \centering
    \label{tab:Stiffness Kresling}
    \begin{tabular}{c c c c }
    \hline
        \textbf{Slit length
(\%of fold line)} & \textbf{FE} & \textbf{Experiment} & \textbf{Relative error (\%)}  \\
        \hline
        0 & 20.14 &16.32 & 18.97\%  \\
        5  & 7.01 & 8.31& 18.54\% \\ 
       10  & 2.48 & 2.96 & 19.35\%  \\ 
        20 & 0.97 & 0.78 & 19.59\%  \\
        \hline
    \end{tabular}
\end{table}

\begin{table}[h]
    \caption{Validation of compressive stiffness (N/mm) - Miura-ori}
    \centering
    \label{tab:Stiffness Miura}
    \begin{tabular}{c c c c }
    \hline
        \textbf{Slit length
(\%of fold line)} & \textbf{FE} & \textbf{Experiment} & \textbf{Relative error(\%)}  \\
        \hline
        0 & 108.07 &47.59 & 55.96\%  \\
        5  & 12.08 & 11.49 & 4.88\% \\ 
       10  & 9.16 & 7.49 & 18.23\%  \\ 
        20 & 2.87 & 2.79 & 2.79\%  \\
        \hline
    \end{tabular}
\end{table}

\clearpage
\newpage

\section{Smooth Fold Lines}
\label{AP:Implementation of Smooth Fold Line Model}
\begin{figure}
 	\centering
    \includegraphics[width=0.88\linewidth]{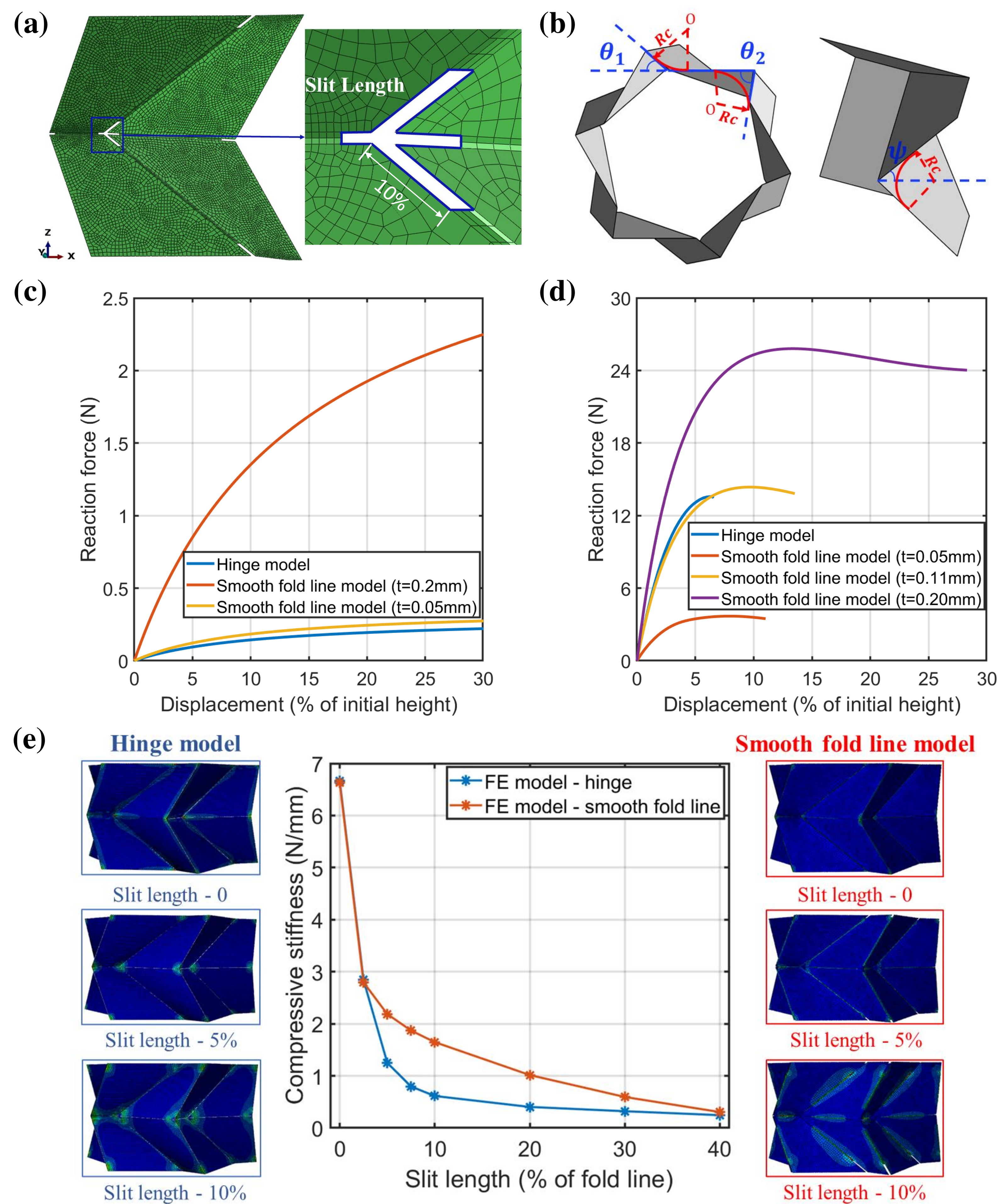}
    \caption{Smooth fold line models ($n=6$, $\phi_1=71^{\circ}$, $\phi_2=41^{\circ}$, $H=39$~mm and $R=65$~mm, bistable), Flexible BC. (a) A unit cell of a finite element model of a Miura-ori bellow with smooth fold lines; the mesh size is $40 \times 40$ elements per facet and 10\% of the fold line is cut open at the vertex, which is highlighted in blue. (b) Definition of smooth fold line models based on straight fold line models; $R_c$ and $O$ are radius and the centre of the smooth creases, and the smooth creases are highlighted in red. (c) Force-displacement curves of hinge models and smooth fold line models with slit length = 0\% of the fold line for a single rigid-foldable unit cell. The thickness at the crease is reduced to $0.05$~mm to match the response of the hinge model. (d) One layer model; the thickness of the smooth crease is reduced to $0.11$~mm to match the response of the hinge model. (e) Compressive stiffness of hinge models and smooth fold line models with different slit lengths; the deformed configurations show the von Mises stress, with warmer colour indicating higher values.}
   \label{fig:Smooth fold line models}
\end{figure}

\noindent The modelling idealisation of a zero-width fold line with linear torsion spring was refined by introducing a smooth fold line: a cylindrical section joins the adjacent facets. A unit cell with a smooth fold line is shown in Fig.~\ref{fig:Smooth fold line models}(a), where the radius of crease $R_c$ is 0.85~mm. The width of the smooth crease $W_c$ is dependent on the radius $R_c$ (see Figure~\ref{fig:Smooth fold line models}(b)) , which can be calculated as:
\begin{equation}
  W_c = (\pi - \theta_1)R_c \quad
   \text{or} \quad W_c = \theta_2 R_c
   \label{}
\end{equation}

In origami structures, the bending stiffness at the creases is lower than that of the facets to facilitate localised folding. This could be modelled by decreasing the material modulus at the crease \citep{woo2008effective}. Here, we achieve this by locally reducing the material thickness along the crease. One unit cell in the bellow (geometric and material parameters are the same as that in the main letter) without slits is taken to calibrate the thickness of creases to match the result of the hinge model; note that this single unit cell is rigid-foldable as the objective is to match the fold stiffness. The boundary condition is the same as that applied in Appendix~\ref{AP:planar-Miura-ori}. It is found out that the thickness of creases should be reduced from 0.2~mm to 0.05~mm to match the result (Figure~\ref{fig:Smooth fold line models}(c)). However, for one layer of a Miura-ori bellow without slits, applying crease thickness $t = 0.05$~mm will underestimate the stiffness of the whole model. The discrepancy between the thickness can be explained by the introduction of crease shearing in the one layer model while the unit cell model can be regarded as rigid-foldable. Therefore, the thickness is adjusted to 0.11~mm (Figure~\ref{fig:Smooth fold line models}(d)), and this value is used in the simulation.

As shown in Figure~\ref{fig:Smooth fold line models}(e), the rapid reduction of compressive stiffness as a result of increasing slit length is observed in both the hinge model and smooth fold line model. This indicates that the influence of slit length is robust to different crease modelling methods.

\end{document}